# The Full Pareto Frontier as Kantian Equilibria

Igor Sloev[1 2], Gerasimos Lianos[3]

Multiplicative Kantian equilibrium explains cooperative behavior in social dilemmas without abandoning methodological individualism. However, its outcomes depend critically on the parametrization of the strategy space — the property of strategic non-equivalence. We investigate what fraction of the Pareto frontier can be attained by varying the strategy space. We show that the set of achievable Kantian equilibria is the *entire* Pareto frontier: for any interior Pareto-efficient point there exists a shift of coordinates — imposing lower bounds on actions — that makes it a Multiplicative Kantian equilibrium. The proof is constructive and relies on a intuitive geometric property: moving the origin to a point on the common tangent to players' indifference curves. This result separates the problem of efficiency from the problem of fairness, allowing any normative criterion to be implemented without loss of Pareto optimality.

[1]School of Computational Social Sciences, European University at St. Petersburg, Russia, ORCID: 0000-0003-0286-1034, sia@eu.spb.ru.

[2]Corresponding author: 6/1A Gagarinskaya st., St. Petersburg 191187 Russia, sia@eu.spb.ru.

[3]Higher Colleges of Technology, Abu Dhabi College, Baniyas, Abu Dhabi, UAE, glianos@hct.ac.ae

## 1. Introduction

The standard result of noncooperative game theory is that in social dilemmas—games with public goods, common-pool resources, or externalities—individually rational behavior, described by Nash equilibrium, leads to collectively irrational outcomes (Hardin, 1968). Yet numerous empirical studies document substantially higher levels of cooperation than Nash predicts, calling for formal models that can explain cooperative behavior without abandoning methodological individualism (Ostrom, 1990).

One of the most influential theoretical responses to this challenge is the concept of Multiplicative Kantian Equilibrium (MKE), introduced by J. Roemer (Roemer, 2010; 2019). An attractive property of MKE is its Pareto efficiency (Roemer, 2010; 2019).

The MKE concept has been applied across several areas of economics: international trade and tax competition (Eichner and Pethig, 2020), environmental economics (Grafton, Kompas and Long, 2017; Ulph and Ulph, 2024; Kurata and Long, 2025), industrial organization (Donduran and Ünveren, 2021; Nakamura, 2023), public goods and taxation (Aronsson and Johansson Stenman, 2007), and health economics (De Donder et al., 2025). This indicates that Kantian optimization is increasingly viewed not merely as a normative ideal, but as a realistic model of behavior across a wide spectrum of economic and social interactions.

An important feature of MKE is strategic non-equivalence: players' payoffs in an MKE depend on the parametrization of the strategy space (Roemer, 2020; Sher, 2020; Sloev and Lianos, 2026). Within the same game, different parametrizations can lead to non-existence of MKE, a unique MKE, or multiple MKE outcomes. Thus, how strategies are formulated critically affects the MKE outcome.

In view of strategic non-equivalence, the following question naturally arises: since any MKE is always Pareto-efficient, what fraction of the Pareto frontier can be reached through Kantian behavior under an appropriate parametrization of the strategy space? The aim of this paper is to show that this subset is the entire Pareto frontier. We prove that for any game in the class we consider and any predetermined Pareto-efficient outcome, there exists a parametrization of strategies—namely, introducing lower bounds on players' actions—under which that outcome is an MKE. The proof is constructive: we derive explicit formulas that use only the properties of the payoff functions and the coordinates of the chosen efficient point. Thus, we not only guarantee the existence of suitable institutional constraints but also provide a constructive algorithm for computing them.

This result is particularly relevant when an institutional designer or society as a whole pursues a specific normative goal—for example, the utilitarian maximum of the sum of payoffs, Rawlsian fairness (maximin), or the Nash bargaining solution (Nash, 1950; Kalai and Smorodinsky, 1975). The classical Kantian approach, while guaranteeing efficiency, does not by itself provide the variety of outcomes necessary for a meaningful choice among competing principles of justice.

The result resonates with the Second Welfare Theorem in the completeness of coverage of Pareto optima: in our setting, any efficient allocation can be decentralized as an MKE by appropriately imposing lower bounds. This shifts the attention of researchers and policymakers from the question “Is cooperation possible?” to “Which cooperation do we choose?”, opening the way to incorporating the full spectrum of normative criteria—from utilitarianism to Rawlsianism—into applied models.

## 2. Multiplicative Kantian Equilibrium

Consider a game $G = i = 1, \dots, n; U_i(x); x_i \geq 0$, where *n* is the number of players, $x_i \geq 0$ is the strategy of player $i$, and $U_i(x)$ is the payoff of player $i$ under the strategy profile $x = (x_1, \dots, x_n)$. We assume that each payoff function $U_i$ is concave in all variables and strictly concave in the player's own strategy $x_i$; externalities are unidirectional: $\partial U_i / \partial x_j$ is either positive for all $i$ and all $j \neq i$ or negative for all $i$ and all $j \neq i$. This specification of the game *G* describes such social dilemmas as the tragedy of the commons, voluntary provision of a public good, and Cournot competition.

Following Roemer (2010), a strategy profile $x^* = (x_1^*, \dots, x_n^*)$ is a *Multiplicative Kantian Equilibrium (MKE)* if

$U_i(x^*) \geq U_i(ax^*)$ for all $a \geq 0$ and all $i = 1, \dots, n$.

That is, no player can gain by changing his own strategy provided all other players change their strategies proportionally. (This proportional change of strategies echoes the Kantian categorical imperative: act as you would wish others to act, which gave the equilibrium its name.)

From the definition of MKE it follows that at $x^*$ we must have

$argmax_a U_i(ax^*) = 1$

Differentiating with respect to $a$ and evaluating at $a$ = 1 yields the necessary first-order condition for an MKE:

$$\partial U_i(x^*)/\,\partial x_1\, x_1^* \;+\; \cdots \;+\; \partial U_i(x^*)/\,\partial x_n\, x_n^* \;=\; 0 \text{ for all } i = 1, \dots, n. \qquad (1)$$

Because $U_i$ is concave in all variables, the function $U_i(a\, x^*)$ is concave in $a$; hence condition (1) is also sufficient, and $a$=1 delivers a global maximum.

**Proposition 1** (Roemer, 2010). If externalities are unidirectional, then every strictly positive MKE, $x^* \gg 0$, is a Pareto-efficient outcome.

### 3. The Geometric Structure of MKE and the Full Realizability Theorem

Our first result follows directly from condition (1) and Proposition 1.

**Theorem 1 (Characterization of MKE).** Under the assumptions of game *G*, every MKE is Pareto efficient and satisfies condition (1). Conversely, any interior Pareto-efficient profile that satisfies condition (1) is an MKE.

*Remark*. By concavity of $U_i$, condition (1) alone is sufficient for $x^*$ to be an MKE. However, in the presence of unidirectional externalities every MKE is Pareto efficient (Proposition 1). Thus, condition (ii) is not redundant in the characterization: it highlights that the set of MKE is exactly the subset of the Pareto frontier selected by condition (1).

The practical value of Theorem 1 is that we need not search for solutions of (1) over the entire strategy space; it suffices to examine the set of Pareto-efficient combinations.

We now examine the geometric interpretation of MKE.

Consider case n = 2. Take an arbitrary point $x_p = \left(x_1^p, x_2^p\right)$ on the Pareto frontier. At $x^p$ the agents' indifference curves touch, i.e., they share a common tangent line orthogonal to the gradients $\nabla U_1$ and $\nabla U_2$. Let $U_{1,x1} = \partial \mathrm{U}_1(\mathrm{x}^\mathrm{p})/\,\partial \mathrm{x}_1$ and $U_{1,x2} = \partial U_1(x^p)/\partial x_2$. The equation of this tangent is

$$x_1 = -U_{1,x_2}/U_{1,x_1}\, x_2 + \left(x_1^p + U_{1,x_2}/U_{1,x_1} x_2^p\right). \qquad (2)$$

If this tangent passes through the origin, the vector $x^p$ lies on it and is therefore orthogonal to the gradient — the necessary MKE condition holds: $U_{1,x_1} x_1^p + U_{1,x_2} x_2^p = 0$.

If the tangent does not pass through the origin, one can always find a point $c = (c_1, c_2)$ on it such that $0 < c_1 < x_1^p$ and $0 < c_2 < x_2^p$. In the new coordinates $z_1 = x_1 - c_1$, $z_2 = x_2 - c_2$, the vector $z^p = \left(x_1^p - c_1, x_2^p - c_2\right)$ lies on the tangent and is thus orthogonal to the gradients $\nabla U_i$. Hence, the Pareto-efficient outcome $x^p$ becomes an MKE in the $z$-coordinates.

Consider case n > 2. The argument is analogous. Let $x^p \in R^n_{++}$ be an interior Pareto-efficient point. By Pareto efficiency, the gradients of all players lie in the same $(n-1)$-dimensional plane and satisfy $\sum_i m_i \nabla U_i = 0$for some positive coefficients $m_i$. This guarantees the existence of a common tangent line through $x^p$; its direction vector $v$ is orthogonal to all $\nabla U_i$. Because each $\nabla U_i$ has components of the same sign, the vector $v$ can be chosen so that moving from $x^p$ along $v$ reduces all coordinates; thus one can pick a point $c = x^p - \varepsilon v$ with $\varepsilon > 0$ sufficiently small and $0 < c_i < x_i^p$ for all $i$. After the change of variables $z_i = x_i - c_i$, the point $x^p$ satisfies condition (1) and is therefore an MKE.

The above reasoning constitutes a constructive proof of the following result.

**Theorem 2.** Let $P = (x_1^p, \dots, x_n^p) \gg 0$ be an interior Pareto-efficient point. Then there exist nonnegative numbers $c_1, \dots, c_n$ such that, in the game with strategy space $z_i = x_i - c_i \geq 0$, the profile $z^* = (x_1^p - c_1, \dots, x_n^p - c_n)$ is an MKE. Geometrically, the vector *c* lies on the common tangent line to the indifference surfaces of all players at *P*, with $c_i < x_i^p$ for all $i$.

The constraint $z_i \geq 0$ merely defines the domain of the transformed game; for the chosen Pareto point all $z_i^p = x_i^p - c_i$ are positive by construction, while the outcomes of the original game that lie outside this domain are irrelevant to the realizability of the given efficient outcome.

### 4. Discussion and conclusion

The theorem presented above separates two fundamental questions of collective action: *how* efficiency is achieved and *which* efficiency is realized. Belonging to the Pareto frontier is guaranteed by Kantian optimization of the incremental efforts $z_i = x_i - c_i$; the specific location on that frontier is determined by the choice of the reference vector *c*. This reformulates the classical "efficiency versus fairness" dilemma as a choice among efficient outcomes. In our approach, once Pareto efficiency is guaranteed by Kantian optimization, the remaining question is which point on the Pareto frontier is selected via the reference vector *c*. This gives the social planner a decentralized mechanism compatible with normative criteria ranging from utilitarianism to Rawlsian fairness (Roemer, 2019; Boadway et al, 2007).

From a positive perspective, the model offers an alternative to the traditional dichotomy of "selfishness versus altruism" or "Nash versus Kant." Our result allows for the interpretation that observed cross-country or cross-industry differences in cooperative behavior may reflect not different degrees of adherence to the Kantian principle, but different historically or institutionally determined reference points $c$. For example, firms in different jurisdictions with similar technological opportunities may voluntarily undertake different levels of environmental investment—this need not be an anomaly but a natural manifestation of Kantian coordination relative to different baseline levels (De Donder et al., 2025; Grafton, Kompas & Long, 2017).

Our model abstracts from the process of forming consensus on the reference vector $c$; it shows that *any* agreed-upon reference vector can be decomposed into an efficient equilibrium. However, the very idea of choosing a reference point as a matter of social contract finds support in the theory of justice and in experimental economics: people often find it easier to agree on minimum standards than on final distributions (Rawls, 1971; Kessler & Leider, 2021). In this light, $c$ can be regarded not only as a technical parameter, but also as the outcome of a legitimate procedure.

Thus, with an appropriate choice of lower bounds, multiplicative Kantian equilibrium can decentralize any Pareto-optimal state. This definitively shifts the focus of the research agenda from the question "Is cooperation possible?" to "Which cooperation do we choose?" and opens avenues for studying the processes of forming the vector $c$ at the intersection of game theory, welfare economics, and political economy.